# Integrated Phononic Waveguides in Diamond

Sophie Weiyi Ding[1*], Benjamin Pingault[1,2], Linbo Shao[1,3], Neil Sinclair[1], Bartholomeus Machielse[1,4], Cleaven Chia[1], Smarak Maity[1], Marko Lončar[1*]

[1]John A. Paulson School of Engineering and Applied Sciences, Harvard University, Cambridge, Massachusetts 02138, USA
[2]QuTech & Kavli Institute of Nanoscience, Delft University of Technology, PO Box 5046, 2600 GA Delft, The Netherlands
[3]Bradley Department of Electrical and Computer Engineering, Virginia Tech, Blacksburg, Virginia 24061, USA
[4]AWS Center for Quantum Networking, Boston, MA 02135, USA (work done prior to joining Amazon)
*Correspondence to: wding@g.harvard.edu (S.W.D.); loncar@seas.harvard.edu (M.L.)

**Abstract**

Efficient generation, guiding, and detection of phonons, or mechanical vibrations, are of interest in various fields including radio frequency communication, sensing, and quantum information. Diamond is an important platform for phononics because of the presence of strain-sensitive spin qubits, and its high Young's modulus which allows for low-loss gigahertz devices. We demonstrate a diamond phononic waveguide platform for generating, guiding, and detecting gigahertz-frequency surface acoustic wave (SAW) phonons. We generate SAWs using interdigital transducers integrated on AlN/diamond and observe SAW transmission at 4-5 GHz through both ridge and suspended waveguides, with wavelength-scale cross sections (~1 μm$^2$) to maximize spin-phonon interaction. This work is a crucial step for developing acoustic components for quantum phononic circuits with strain-sensitive color centers in diamond.

**Introduction**

Acoustic devices at microwave frequencies have been studied intensively because of the advantages they offer over their electromagnetic counterparts: smaller wavelengths, and hence reduced sizes, lower crosstalk, and lower losses [1–3]. These properties also make them good candidates for applications in quantum science and technology since they can provide efficient interfaces between microwave signals and qubits. As a result, quantum phononics, or quantum acousto-dynamics (QAD), which relies on phonons as information carriers, has emerged as an active field of study [4–11]. One promising QAD approach uses surface acoustic waves (SAWs) to control and read out qubit states. SAW-based quantum interfaces have been demonstrated for various qubits, including color centers in diamond [12,13] and silicon carbide [14,15], quantum dots [16], and superconducting qubits [17,18].

Among these, the electron spin qubit associated with the negatively charged silicon-vacancy (SiV) color center in diamond has drawn a lot of interest. Diamond, with its high Young's modulus (>1000 GPa), is an ideal platform for gigahertz-frequency phononics; SiV itself is a promising quantum memory [19,20] that can be efficiently interfaced with phonons owing to its high strain susceptibility of ~100 THz [21,22] while giving access to long-lived nuclear spin-based quantum registers [23]. It has been shown that SAW generated with integrated IDTs can control a single SiV electron spin with a high Rabi rate ~30 MHz, as well as a nuclear spin in diamond, using power orders of magnitude lower than for standard microwave control [12]. However, these works are performed in bulk diamond without any waveguiding of the SAW and poor optical collection efficiency, thereby limiting the scalability of this approach and its integrability with other devices. To develop an integrated platform comprising multiple spin qubits with enhanced spin-phonon interaction and to more efficiently guide and detect SAWs**,** integrating IDTs with micro/nanostructures such as phononic waveguides or cavities is a necessary step. This furthermore allows for leveraging recent progress with diamond photonic structures for both on-chip spin-phonon and long-distance spin-photon networking of multiple spins.

Here, we experimentally demonstrate integrated phononic waveguides with a ridge as well as suspended structures in single crystal diamond, which can support phonon modes at high frequencies over 4 GHz. The ridge waveguides, fabricated into AlN-on-diamond, are advantageous for structural stability and strong thermal contact with the diamond substrate. We further develop these structures and fabricate ridge crossing-waveguides and observe a large suppression of the cross-talk between the two intersecting waveguides. To realize better confinement of both phonons and photons in waveguides, we demonstrate suspended AlN/diamond nanobeam waveguides and observe SAW transmission. Our results using integrated phononic waveguides pave the way not only for efficient generation, guiding and detection of SAW on diamond, but also for manipulating spin qubits using SAW with low loss and low power consumption combined with future photonic integration.

**Acoustic ridge waveguide: design and fabrication**

An optical micrograph of the fabricated ridge-waveguide device is shown in Fig. 1(a). Interdigitated transducers (IDTs), patterned on AlN, are used for bidirectional conversion between SAWs and microwaves and are used to generate and detect acoustic waves. We pattern the AlN film to form an acoustic waveguide between the two IDTs: since diamond has a higher speed of sound (12~18 km/s) than AlN (6~10 km/s) [24,25], the traveling wave is bound to the interface between AlN and diamond. Thus, as the width of the AlN region is reduced, the SAW becomes more confined in both AlN and in the diamond beneath.

To design our waveguides, acoustic wave mode simulations have been performed using the finite element method (FEM) in COMSOL Multiphysics. Fig. 1(b) shows the cross-section of fundamental shearing and fundamental Rayleigh modes, along with their normalized displacement in the coupler region. Fig. 1(c) shows the two modes at the ridge waveguide. In Fig. 1(b), the acoustic wavelength is 2.4 μm, which corresponds to shearing and Rayleigh mode frequencies of 3.2 GHz and 4.7 GHz, respectively.

Fabrication of the device begins by sputtering 700 nm of AlN on a single-crystal diamond substrate. The AlN thickness is chosen to maximize the electromechanical coupling efficiency, $k^2$, for a AlN/diamond interface (Supplementary Material). To pattern the 2 μm-wide waveguide, we first define an etch mask in HSQ using electron beam lithography (EBL). Then the AlN layer is etched down using reactive-ion etching (RIE) with argon and chlorine gasses. Finally, the IDTs and the contact pads are written by EBL using PMMA resist, and created by liftoff process after depositing a 100 nm-thick Al layer using electron beam (EB) evaporation.

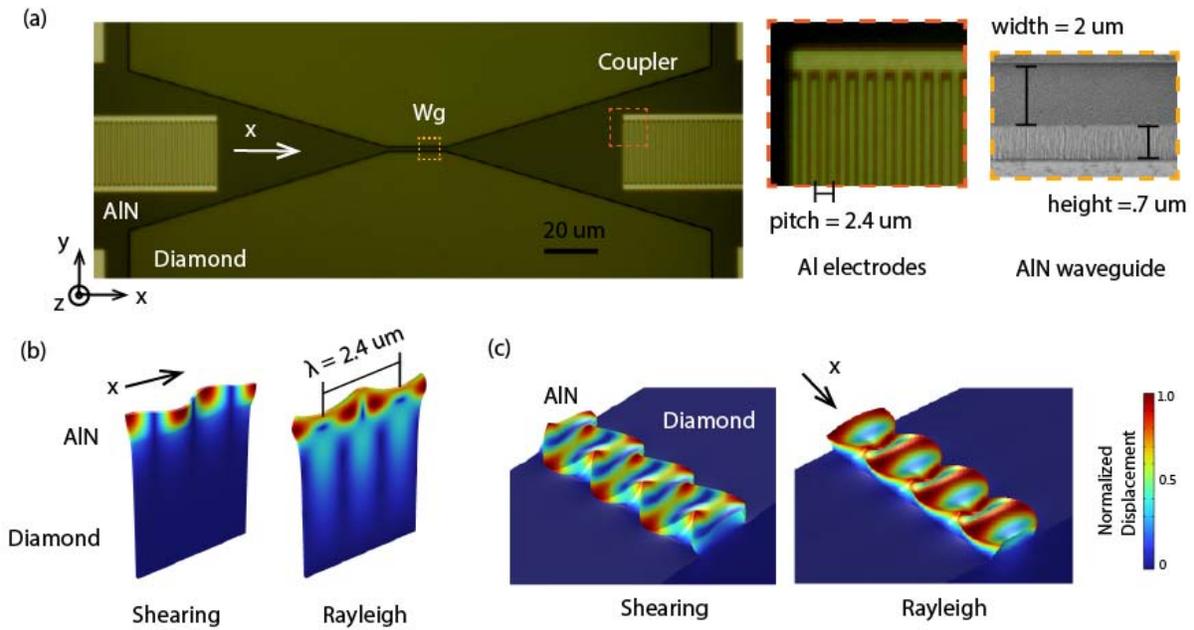

**Fig. 1. A surface acoustic ridge waveguide device using AlN-on-diamond.** (a) Images of the device. The microscope image on the left shows a top-down view of the entire device. The zoomed-in pictures illustrate the detail of the electrodes (microscope image) and the side profile of the waveguide (SEM image, taken at 60 deg tilt). (b) The cross-section of the two fundamental SAW modes, shearing and Rayleigh modes, in the coupler region. An 'x' indicates the propagation direction of the wave. (c) The SAW modes propagate in the waveguide section.

**Acoustic ridge waveguide: characterization**

The device is characterized in the frequency and time domains using the setup shown in the inset of Fig. 2 (b). The measurement reveals three transmission peaks, around 3.2, 4.4, and 7.1 GHz (Fig 2(a)). The first two peaks correspond to fundamental shearing and Rayleigh modes, and are in good agreement with numerical modeling shown in Figure 1(b). The third peak with the highest frequency is a higher order shearing mode. The Rayleigh mode, shown in Fig. 2(b) has the largest transmission because it has the highest $k^2$ among the modes (Supplementary Material). The 3-dB bandwidth of the Rayleigh mode is 10 MHz, which is smaller than the predicted IDT bandwidth. We attribute this to loss and to the presence of the waveguide that has lower effective acoustic velocity compared to the bulk. The maximum transmission ($|S_{21}|$) of the Rayleigh mode is measured to be -22.4 dB, which is comparable to previous ridge phononic waveguides in low-loss thin fim platforms [26–28]. The corresponding reflection ($|S_{11}|$) is -4.0 dB, which is close to the calculated reflection of -5.2 dB, due to impedance mismatch (Supplementary Material).

The time domain measurement is performed using the setup in the inset of Fig. 2(c). We excited our waveguide with a 5-ns and a 20-ns pulse at 4.45 GHz, which correspond to a pulse width of 200 MHz and 50 MHz in frequency, respectively. The measured transmission shown in Fig. 2(c) has several important features. The first pulse (light blue) observed in the time-domain corresponds to direct crosstalk between microwave cables used to connect two IDTs, since microwave signals in free space propagate faster than acoustic waves. This pulse has the profile of the input microwave pulse with vestigial noise. The pulse that is delayed by ~15 ns from the crosstalk (dark blue, pulse 1) is the SAW transmission. The delay time corresponds to the distance between the IDTs divided by the acoustic velocity. The SAW transmission is broadened compared to the input pulse due to dispersion imparted by the waveguide. The second pulse (dark blue, pulse 2) is attributed to the SAW reflected by IDTs and is delayed by ~3x15 ns, which is the round-trip time. The reflection can be mitigated by an asymmetric design, or can be enhanced through phononic structures to make high quality factor SAW cavities[14,29]. This time-domain measurement confirms the acoustic nature of the signal that can be used to drive SiV spin qubits mechanically.

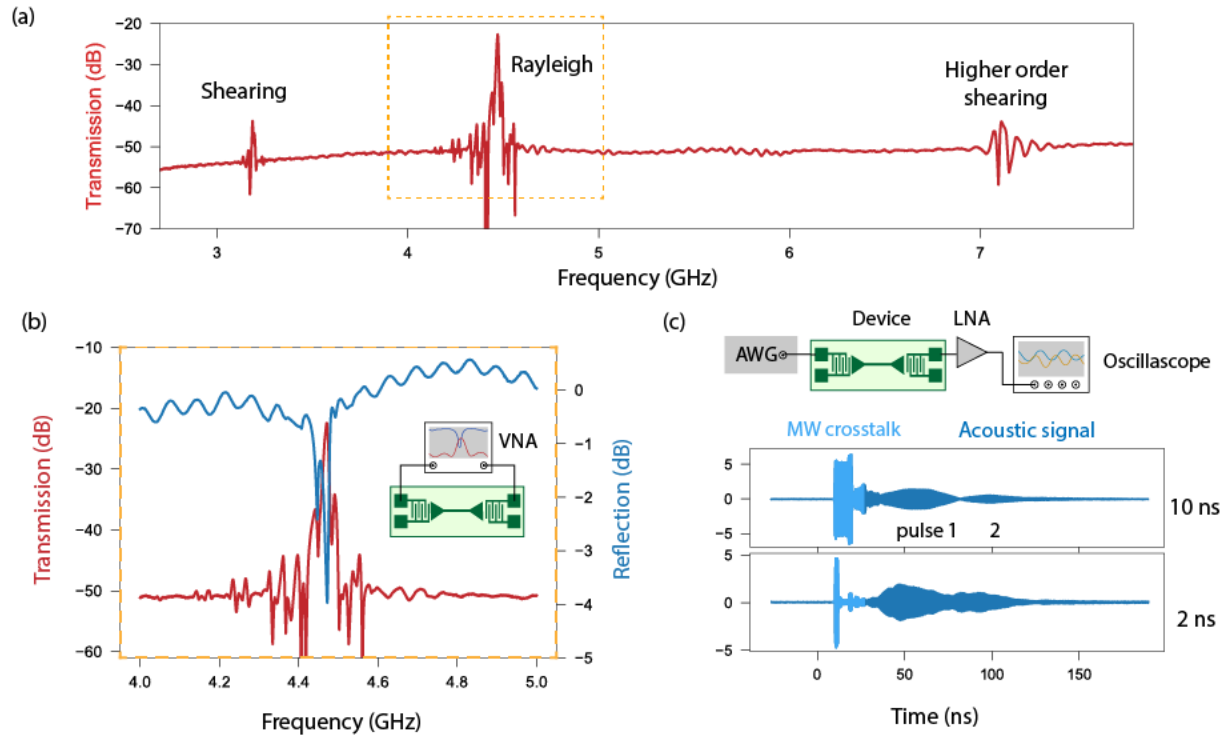

**Fig. 2. Characterization of an AlN-on-diamond acoustic waveguide**. (a) Frequency domain measurement of the waveguide. The scan shows three transmission peaks, corresponding to the lowest frequency modes we have simulated. The Rayleigh mode has the highest transmission. (b) Frequency domain measurement of the Rayleigh mode. The peak frequency is 4.45 GHz and the 3-dB bandwidth is 10 MHz. The inset depicts the measurement setup. VNA: vector network analyzer. (c) Time domain measurement setup and response for the Rayleigh mode in arbitrary units. The light blue pulse that arrives first is the microwave crosstalk, pulse 1 is the transmitted SAW, and pulse 2 is the reflected SAW. AWG: arbitrary waveform generator, LNA: low noise amplifier.

Based on the realized ridge waveguides, we also demonstrate low-crosstalk waveguide crossings for efficient routing of acoustic waves. Figure 3 (a) shows the structure of the waveguide crossings consisting of two perpendicular waveguides with IDTs at their ends. We first simulate the transmission property in the waveguide crossings by exciting one of the ports with the Rayleigh waveguide mode (Fig. 3 (b)) while monitoring SAW transmission into the remaining three waveguide ports. By comparing the energy transmitted into through- and cross-ports, we estimate crosstalk to be -15.8 dB. We also experimentally characterize fabricated devices and measure a maximum transmission of - 30.6 dB and -46.9 dB into through- and cross-ports, respectively, resulting in a crosstalk of -16.3 dB (Fig. 3 (c)). This is in good agreement with the simulations. The transmission into the through-port is lower than that in a simple waveguide geometry shown in Fig. 2, because of deflection and scattering loss at the crossing points.

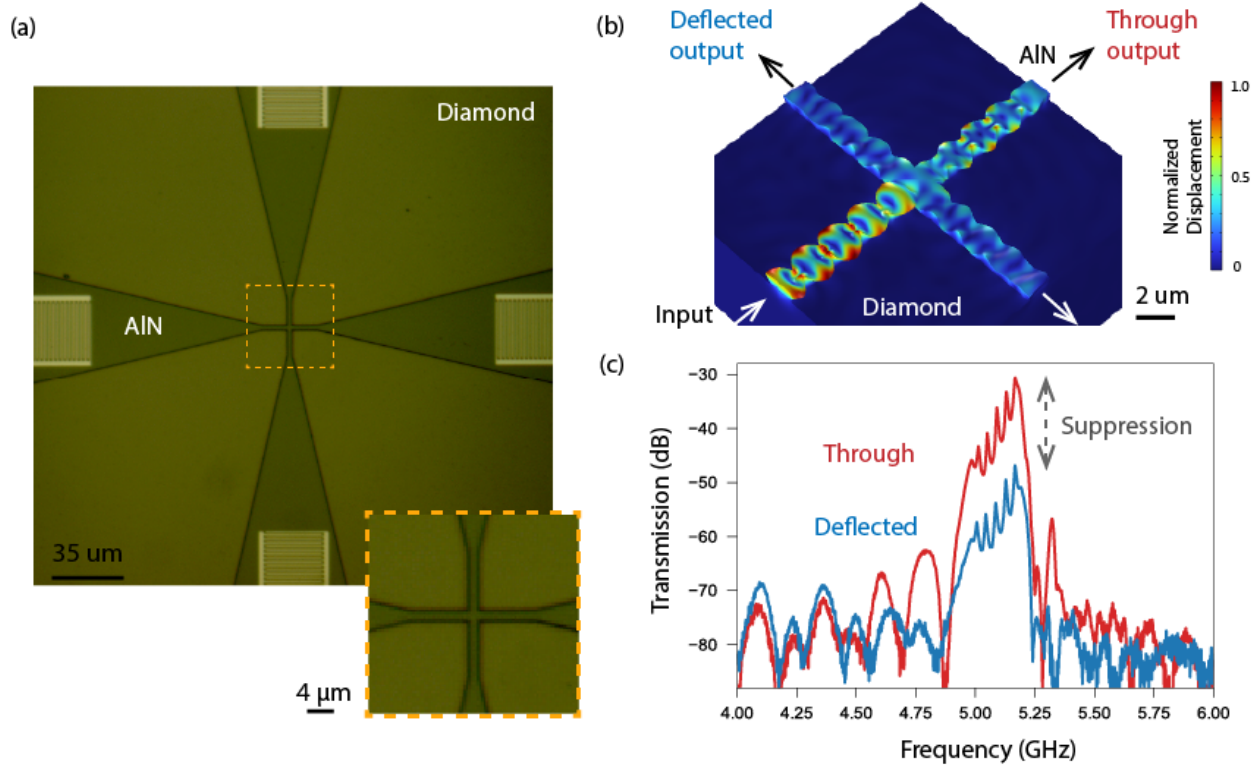

**Fig. 3. Acoustic waveguide crossing and crosstalk measurement**. (a) Optical microscope image of the cross-waveguide device used to evaluate crosstalk. There is an IDT placed at the end of each waveguide for excitation and collection of SAWs. The inset shows the zoom-in into the cross. (b) Simulation of the acoustic Rayleigh mode propagating through the waveguide crossing. (c) Transmission spectra of SAWs of the through and one of the deflected ports of the cross, indicating a maximum crosstalk of -16.3 dB.

### Free-standing Acoustic Waveguides

One of the shortcomings of the bulk approach discussed above is weak confinement and scattering into the substrate, which can result in acoustic losses. Furthermore, these AlN-on-diamond devices do not support confined optical modes (optical waveguides or cavities) that could be used to e.g. efficiently collect photons emitted by SiV [20]. For more general acousto-optical applications, we also would like to explore tighter confinement of both optical and acoustic modes [30,31]. To address these points, we consider suspended acoustic waveguides (Fig. 4).

Our device consists of an IDT that is used to excite a SAW, followed by a linear taper used to focus the acoustic energy into the free-standing region and excite its acoustic modes. The dispersion diagram for the modes supported by the suspended waveguide are shown in Fig. 4(b). The two fundamental modes are shown in blue and red, corresponding to flexural versions of the shearing and Rayleigh SAW modes in the ridge waveguide, respectively. Acoustic energy transmitted through the free-standing section is then collected using another tapered section and finally detected using an IDT. The coupler region is designed in a similar fashion as in the case of Fig. 2 and 3.

Device fabrication starts with the deposition of AlN. Next, the electrodes are defined by electron-beam lithography and liftoff using Cr/Au (100 nm). The coupler and waveguide patterns are defined using EBL with HSQ resist. Next, RIE with $O_2$ plasma is used to etch the pattern into the diamond, followed by quasi-isotropic etching to undercut the waveguide and release it from the diamond substrate [32,33]. The suspended section is a 1-μm waveguide, consisting of 700 nm of AlN on 200 nm of diamond (Fig. 4(a)). This thickness of the diamond is suitable for implantation and study of defects for quantum experiments. We note that tapering between the coupling region and the free-standing section is three-dimensional: both laterally and orthogonally from the diamond surface.

We perform scattering-parameter measurements and observe a mode at 4.18 GHz with a 3-dB bandwidth of 7 MHz. The transmission and reflection are measured to be -50 dB and -1.8 dB at the peak, respectively, comparable to previous only-AlN-suspended-phononic-waveguide works [30,34]. We found that the resonance frequency is different from the ridge waveguide of the same IDT pitch (4.45 GHz). This is due to the suspended waveguide and the use of Cr/Au electrodes, which are heavier. Cr/Au is chosen over Al for the suspended structure to avoid etching by HF in the step to remove the HSQ etch mask. The lower transmission compared to the ridge waveguide device is due to a smaller modal overlap of the SAW modes in the coupler region and the suspended waveguide mode. The roughness on the bottom of the waveguide introduced by the quasi-isotropic etch also results in reflection and scattering into other waveguide modes, thus reducing the overall transmission. These challenges are associated with other suspended platforms as well, particularly at the μm-wavelength scales, as active research is being conducted on improving and harnessing their unique properties [30,31,35].

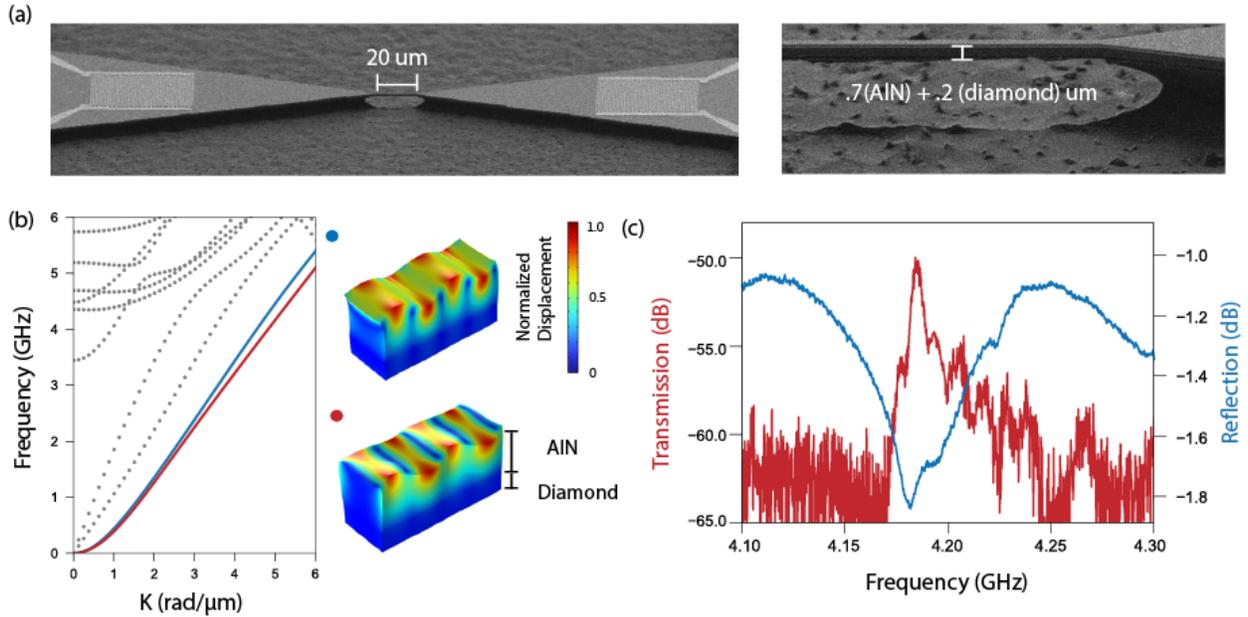

**Fig. 4. Suspended acoustic AlN-on-diamond waveguide device and measurement**. (a) SEM images (taken at 60 deg tilt) of a suspended acoustic waveguide. The waveguide is excited using IDTs defined on non-suspended regions. The zoomed-in figure shows the details of the taper and the composition of the waveguide: 700 nm of AlN on 200 nm of diamond. The lighter top region is an AlN chamfer from mask erosion. The bottom darker sleeve is AlN and diamond. (b) Dispersion diagram for acoustic modes supported by the suspended waveguide. The red and blue lines correspond to the flexural versions of shearing and Rayleigh modes, respectively, of the ridge waveguide. (c) Frequency-domain transmission spectra of the fundamental modes showing a maximum transmission and reflection of -50 dB and -1.8 dB, respectively.

In this case, to overcome the challenges associated with injecting SAWs from the wide section into the suspended narrow waveguide, we consider the structure shown in Fig. 5. Here, IDTs are defined directly on the free-standing waveguide and can be used to excite the Rayleigh-like mode of the suspended region (Fig. 5(a)), and can be eventually coupled to a phononic crystal for enhanced spin-phonon interaction (Fig. 5(b)). The benefit is less injection loss, more efficient coupling, but the challenge is impedance matching due to the reduced capacitance of narrow IDTs, which is fundamentally limited by the $k^2$ of this material stack. An example fabricated device is shown in Fig. 5(c).

To enable such a geometry, we can use materials with higher electromechanical coupling (e.g lithium niobate and Sc-AlN) as the piezoelectric layer, or implement an external impedance matching circuits for better impedance matching [36]. Additionally, we can fabricate the entire device on a suspended diamond thin film to reduce loss in tapering and allow for larger capacitance [37].

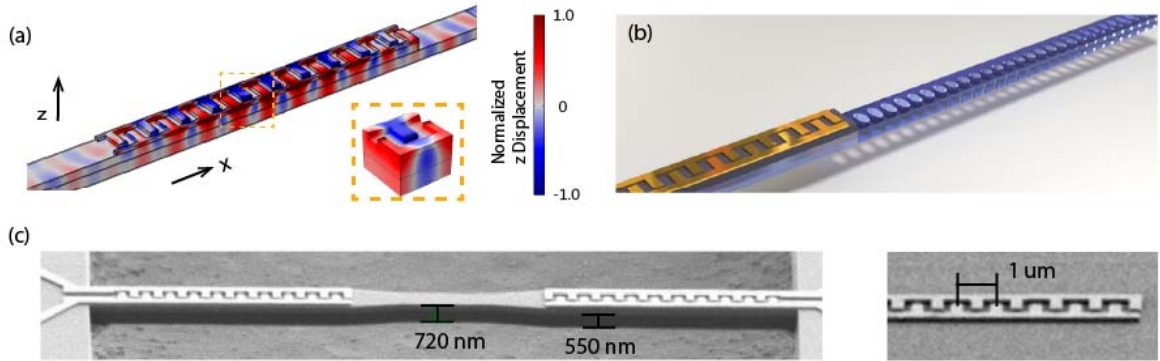

**Fig. 5. Fully suspended acoustic AlN-on-diamond waveguide device.** (a) Simulation of the mode excited by the IDTs. The frequency of the mode is 6.15 GHz, and the wavelength is 1 μm. (b) A rendered image of the suspended IDT coupled to a mechanical resonator. (c) An SEM image (taken at 60 deg tilt) of suspended IDT couplers and a suspended waveguide. The zoomed-in figure shows the metal fingers after the liftoff before the etch.

**Spin-phonon coupling between an SiV and the acoustic waveguide:**

The mode frequencies above are designed to match an SiV spin-qubit splitting of about 5 GHz (Fig 6(a)), that can be readily accessed with an applied magnetic field and is typically used for qubit control [20,38]. Based on the simulated waveguide modes and transmission measurements, we can calculate the spin-phonon coupling rate between the fabricated waveguide and an SiV. The acoustic mode is evanescently coupled through the generated strain to an SiV implanted 50-100 nm below the diamond surface underneath the waveguide.

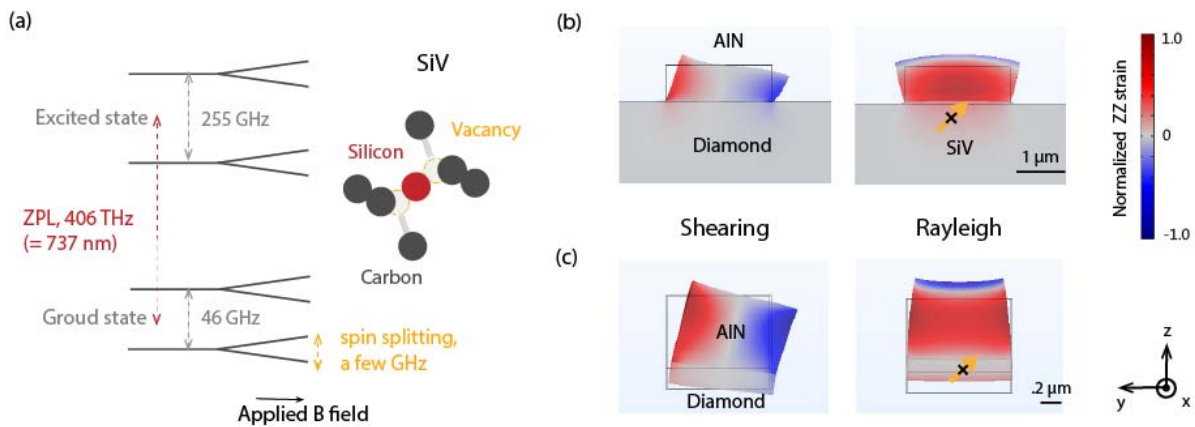

**Fig. 6. SiV and the strain profile of the ridge and suspended waveguide mode** (a) SiV energy level diagram and its atomic structure. The spacings between the levels are not drawn to scale. The spin level of interest is tunable by an applied magnetic field, usually around 3~10 GHz. ZPL: zero phonon line. (b) The cross-section profile of the zz strain in the ridge waveguide for shearing and Rayleigh modes. ZZ strain often contributes the most to the coupling with the given orientations of the SiV spin in diamond. The values are normalized to the maximum zz strain for visualization. The displacement of the mode is shown by the exaggerated deformation. In both cases, the SAW is guided by the waveguide, propagating in the x direction. In the Rayleigh case, strain can be more effectively delivered to the SiV near the surface. (c) The cross-section profile of the zz strain in the suspended waveguide region for shearing and Rayleigh mode.

The spin-phonon coupling rate is given by: $g = \Sigma_{ij} d_{ij} \epsilon_{ij}$, where i,j are indices for the coordinate axes, $\epsilon$ is the strain experienced by the SiV produced by a single phonon, d is the spin level strain susceptibility. The strain profile is obtained from the waveguide simulation and the susceptibility has been measured in previous work [21]. For the ridge case, the single phonon strain is on the order of $10^{-10} \sim 10^{-12}$, with $\epsilon_{zz} = 5.0 \times 10^{-10}$. The susceptibility is ~100 THz/strain for the spin levels of the SiV, depending on the symmetry of the strain and also the spin level splitting, which can be tuned by an external magnetic field [21]. For the ridged waveguide (Fig. 6(b)), the calculated value for a single phonon Rabi rate is 12~15 kHz for the Rayleigh mode; for the suspended waveguide (Fig. 6(c)), the rate is 15~20 kHz.

To compare the rates to previous SAW works [12,23], we can calculate the Rabi frequency while the SiV is driven in the waveguide, using the measured values for transmission. For 1mW input power, with empirical external losses taken into consideration (10 dB per port), the projected Rabi frequency is 5.5 GHz for the ridge structure, which is more than two orders of magnitude higher than previous works [12,23]. In practice, this means we can drive the SiV efficiently at even lower powers. The calculation details are in the Supplementary Material.

**Conclusion and outlook**

We present the fabrication and measurements of a GHz-phononic circuit on diamond for applications in quantum acoustic dynamics (QAD). To improve upon the current bulk diamond platform, we demonstrate the fabrication of both ridge and suspended phononic waveguides, for effective guiding and detection of coherent phonons, as well as for enhanced spin-phonon coupling, two orders of magnitude higher than previous works.

This demonstration allows for the injection and collection of phonons from other diamond structures with state-of-the-art phononic properties, such as diamond cantilevers and optomechanical crystals [39,40]. With the platform improvements and the prospect of combining such devices with SiVs, we hope to open up new opportunities for exploration of the physics of spin-phonon coupling and its applications.

**Acknowledgement**
This work is supported by National Science Foundation (NSF) Science and Technology Center for Quantum Materials (DMR-1231319), NSF Engineering Research Center for Quantum Networks (EEC-1941583), Office of Naval Research (N00014-20-1-2425), Air Force Office of Scientific Research (AFOSR) MURI on Quantum Phononics, and AFOSR (W911NF-23-1-0235). B.P. acknowledges funding from the European Union's Horizon 2020 research and innovation programme under the Marie Skłodowska-Curie Grant Agreement No. 840968.

The views and conclusions contained in this document are those of the authors and should not be interpreted as representing the official policies, either expressed or implied, of the Army Research Office or the U.S. Government. The U.S. Government is authorized to reproduce and distribute reprints for Government purposes notwithstanding any copyright notation herein.

**Appendix A: IDT design**

In order to achieve an efficient acoustic-spin interface, it is important to optimize the geometry of the IDTs used. For example, a large IDT bandwidth (BW) is needed to allow for fast control of SiV spins. This in turn limits the number of IDT pairs ($N$), that can be used, since $BW \sim f_0/N$, where $f_0$ is the operating frequency. However, larger $N$ is needed to match the impedance ($Z$) of the IDT to the standard $50\ \Omega$, thus minimizing the reflection of microwave signals used to excite SAWs. In our design, we chose $N = 30$, which which aims at BW~150 MHz and $Z$~$170\ \Omega$. Notice, because of the symmetric nature of the device, the theoretical maximum transmission would be -6 dB, since each set of the IDTs has a loss of 3 dB. With insertion losses from the tapering and other sources, the measured value is always lower.

The impedance of an IDT is theoretically determined by the real part of its inverse admittance G, which is given by $G = 8k^2 c_s w f_0 N^2$, where $k^2$ is the electromechanical coupling coefficient, $c_s$ is the capacitance per length, $w$ is the IDT width, $f_0$ is the SAW central frequency, and $N$ is the number of fingers of the IDT. We experimentally determined that $N = 30$ yields the highest microwave transmission in our device geometry. In practice, the best N is usually less than that given by the impedance matching condition because of loss related to the SAW generation. This geometry yields an impedance of around 170 Ohms with the above expression, which gives a reflection signal of -5.2 dB. In our case, because of the low $k^2$ of thin film AlN, combined with the need to taper into a narrow waveguide (which limits the width and thus the geometry/total area of the IDT), an exact 50-Ohm impedance matching is challenging to achieve.

**Appendix B: $k^2$ simulation**

The electromechanical coupling coefficient is calculated from the phase velocity difference between the cases where the boundary condition at the IDT coupler surface is electrically free or shorted:

$$k^2 = 2(v_{free} - v_{shorted})/v_{free}.$$

A large $k^2$ means higher conversion efficiency between electrical and acoustic energy in a piezoelectric material. The following phase velocity and $k^2$ simulations are performed using COMSOL. Mode 1 and 2 correspond to the shearing and Rayleigh mode presented in the main text, and Rayleigh has the highest transmission because it has the highest $k^2$ at the given $h/\lambda$.

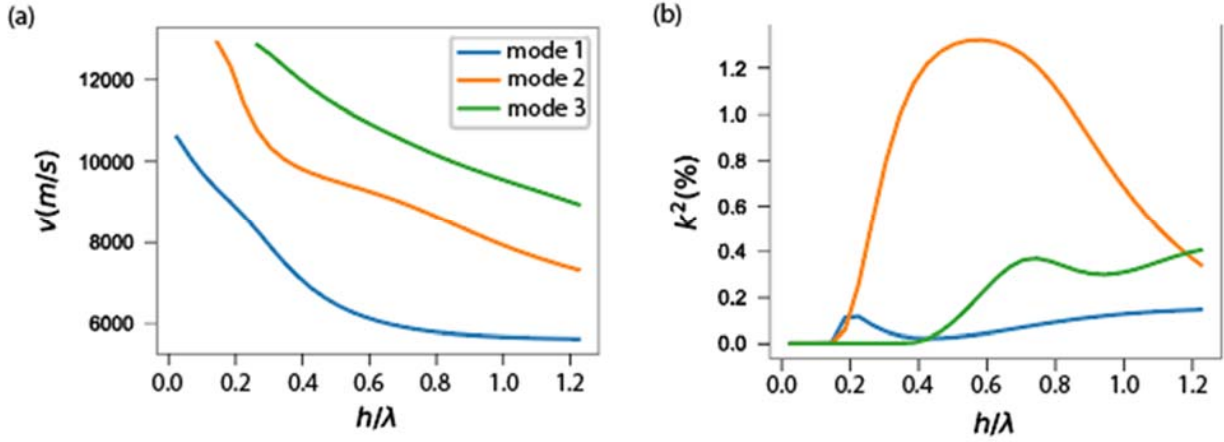

Fig s1. (a) the phase velocity of the AlN/diamond SAW (free) as a function of AlN thickness and wavelength ratio. (b) the $k^2$ of the AlN/diamond SAW as a function of AlN thickness and wavelength ratio. Mode 1 is Rayleigh and has the highest $k^2$, where $h \approx 0.55\lambda$ and $k^2 \approx 1.2\%$.

## Appendix C: Calculating single spin-phonon coupling and Rabi rate

The simulation produces the strain at a selected location and the total amount of energy in the simulation domain. We can obtain the single phonon Rabi rate: $g = strain \times d/\sqrt{n}$, where d is the strain susceptibility of the spin, normalized by the number of phonons: $n = E_{total}/hf$, with frequency f = 4.5 GHz in the ridge case. The susceptibility is ~100 THz/strain for the spin levels of the SiV, depending on the symmetry of the strain [21] and also the spin level splitting, which can be tuned by an external magnetic field [21]. We calculate the coupling rate for a [111] SiV 100 nm below the diamond surface, where the SiV is in the middle of the waveguide Rayleigh mode, as shown in Fig. 6 (b). The single phonon Rabi is calculated to be $g = 12$~$15$ $kHz$. Similarly for the suspended mode, the single phonon coupling rate to the spin is $g = 15$~$20$ $kHz$ for the Rayleigh mode. This coupling is larger because the mode is more confined than for the ridge waveguide.

To compare to previous SAW works [12,23], the figure of merit is how large the Rabi rate can be while the SiV is driven with a small input power. Assume an input power of 1 mW and a tranmission of -20 dB for the ridge waveguide, as measured. We add another -10 dB per port insertion loss from wire-bonding and other connection losses for a realistic setup. The power of a single phonon is: $p_0 = hf/t_p = 5.96 \times 10^{-17}$ W, where $t_p$, the duration of the phonon, is 50 ns and $f = 4.5$ $GHz$. Assume 1% ($k^2$) of the power delivered to the waveguide is converted to the phonons, so the number of phonons in the waveguide is $n = 1.68 \times 10^{11}$, and the corresponding Rabi rate is $\sqrt{n}g = 5.5$ $GHz$, which is more than two orders of magnitude faster than experimentally measured values in previous SAW works with the same power. The Rabi rate scales with the square root of the input power. Therefore, in practice, driving the SiV in this waveguide geometry requires much less power than 1 mW.